\documentclass[acmtog,screen]{acmart}

\usepackage{xspace}
\usepackage{algorithm}
\usepackage{algorithmic}

\newcommand*{\Reals}{\ensuremath{\mathbb{R}}}

\newcommand{\calT}{\mathcal{T}}
\newcommand{\calC}{\mathcal{C}}

\newcommand{\KMG}{\hat{G}_{KM}}

\newcommand{\II}{\ensuremath{\mathbb{I}}}

\newcommand{\pkg}[1]{\textbf{#1}\xspace}
\newcommand{\Rstats}{\textsf{R}\xspace}

\AtBeginDocument{%
  \providecommand\BibTeX{{%
    \normalfont B\kern-0.5em{\scshape i\kern-0.25em b}\kern-0.8em\TeX}}}

\setcopyright{acmcopyright}
\copyrightyear{2022}
\acmYear{2022}
\acmDOI{XXXXXXX.XXXXXXX}

\acmConference[DSHealth 2022]{Workshop on Applied Data Science for Healthcare, DSHealth 2022}{August 14, 2022}{Washington DC}
\acmSubmissionID{9095}

\begin{document}

\title{Flexible Group Fairness Metrics for Survival Analysis}

\author{Raphael Sonabend}
\authornote{These authors contributed equally to this research.}
\email{raphaelsonabend@gmail.com}
\orcid{0000-0001-9225-4654}
\affiliation{%
  \institution{Technische Universität Kaiserslautern}
  \streetaddress{Gottlieb-Daimler-Straße 47}
  \city{Kaiserslautern}
  \country{Germany}
  \postcode{67663}
}
\affiliation{%
  \institution{University of Cambridge}
  \streetaddress{Engineering Department, Trumpington Street}
  \city{Cambridge}
  \country{UK}
  \postcode{CB2 1PZ}
}
\affiliation{%
  \institution{Imperial College London}
  \streetaddress{School of Public Health}
  \city{London}
  \country{UK}
  \postcode{W2 1PG}
}

\author{Florian Pfisterer}
\authornotemark[1]
\email{pfistererf@googlemail.com}
\orcid{0000-0001-8867-762X}
\affiliation{%
  \institution{Ludwig-Maximilians-Universität München}
  \streetaddress{Ludwigstr. 33}
  \city{München}
  \country{Germany}
  \postcode{80539}
}

\author{Alan Mishler}
\email{alan.mishler@jpmchase.com}
\orcid{0000-0002-7654-208X}
\affiliation{%
  \institution{J.P. Morgan AI Research}
  \streetaddress{383 Madison Ave.}
  \city{New York}
  \state{NY}
  \country{USA}
  \postcode{10001}
}

\author{Moritz Schauer}
\email{smoritz@chalmers.se}
\orcid{0000-0003-3310-7915}
\affiliation{%
  \institution{Chalmers Technical University and University of Gothenburg}
  \streetaddress{Chalmers Tv\"argata 3}
  \city{Gothenburg}
  \country{Sweden}
  \postcode{41296}
}

\author{Lukas Burk}
\email{Lukas.Burk@stat.uni-muenchen.de}
\orcid{0000-0001-7528-3795}
\affiliation{%
  \institution{Ludwig-Maximilians-Universit\"at M\"unchen}
  \streetaddress{Ludwigstr. 33}
  \city{München}
  \country{Germany}
  \postcode{80539}
}
\affiliation{%
  \institution{Leibniz Institute for Prevention Research and Epidemiology - BIPS GmbH}
  \streetaddress{Achterstr. 30}
  \city{Bremen}
  \country{Germany}
  \postcode{28359}
}

\author{Sumantrak Mukherjee}
\email{f20190413@pilani.bits-pilani.ac.in}
\affiliation{%
  \institution{Birla Institute of Technology and Science}
  \streetaddress{Vidya Vihar}
  \city{Pilani}
  \state{Rajasthan}
  \country{India}
  \postcode{333031}
}

\author{Sebastian Vollmer}
\email{Sebastian.Vollmer@dfki.de}
\affiliation{%
  \institution{Deutsches Forschungszentrum für Künstliche Intelligenz (DFKI)}
  \streetaddress{Trippstadter Str. 122}
  \city{Kaiserslautern}
  \country{Germany}
  \postcode{67663}
}
\affiliation{%
  \institution{Technische Universität Kaiserslautern}
  \streetaddress{Gottlieb-Daimler-Straße 47}
  \city{Kaiserslautern}
  \country{Germany}
  \postcode{67663}
}
\affiliation{%
  \institution{University of Warwick}
  \streetaddress{Zeeman Building}
  \city{Coventry}
  \country{UK}
  \postcode{CV4 7AL}
}

\renewcommand{\shortauthors}{Sonabend, et al.}

\begin{abstract}
   Algorithmic fairness is an increasingly important field concerned with detecting and mitigating biases in machine learning models. There has been a wealth of literature for algorithmic fairness in regression and classification however there has been little exploration of the field for survival analysis. Survival analysis is the prediction task in which one attempts to predict the probability of an event occurring over time. Survival predictions are particularly important in sensitive settings such as when utilising machine learning for diagnosis and prognosis of patients. In this paper we explore how to utilise existing survival metrics to measure bias with group fairness metrics. We explore this in an empirical experiment with 29 survival datasets and 8 measures. We find that measures of discrimination are able to capture bias well whereas there is less clarity with measures of calibration and scoring rules. We suggest further areas for research including prediction-based fairness metrics for distribution predictions.
\end{abstract}



\maketitle

\section{Introduction}
The use of machine learning (ML) models, especially in the context of clinical decision making \cite{topol2019high} can lead to, or exacerbate, disparities in health outcomes for marginalized populations \cite{nordling2019fairer,obermeyer2019dissecting, vollmer2020machine}. This can arise due to multiple reasons, such as access to different standards of care \cite{bailey2017structural}, historical inequity \cite{chen2020treating, hall2015implicit} or under-representation in data collection \cite{buolamwini2018gender, larrazabal2020gender}. This can lead to models with differing (predictive) effectiveness depending on the subpopulation or models that perpetuate historical injustices if they are subsequently used to inform medical decisions \cite{veinot2018good}. The goal of algorithmic fairness is to detect and mitigate such biases \cite{mehrabi, barocas-hardt-narayanan}. This has been discussed in great detail in classification and regression settings \cite{rajkomar2018ensuring, mehrabi, Pfohl2021-xd}, however very little discussion exists for survival analysis. This is problematic given the sensitive nature of survival predictions. For example, hospitals with insufficient resources may require survival models to accurately and fairly rank patient outcome risks \cite{ryan2020mortality, liang2020early, sprung2020adult}. It is crucial that algorithmic fairness is considered in the survival setting. Zhang and Weiss \cite{Zhang2022} have begun exploring debiasing methods for survival analysis, however only a limited number of measures are considered. In this article we examine whether existing metrics that are used in survival analysis can be adapted to detect unfairness in survival models. The code required to reproduce the results in this paper is publicly available at \url{https://github.com/Vollmer-Lab/survival_fairness}. For a comprehensive overview to fairness and survival analysis we recommend Mitchell et al. (2021) \cite{Mitchell2021} and Wang et al. (2019) \cite{Wang2019} respectively.

\section{Related Work}

\subsection{Fairness metrics}
Many notions of fairness exist, including individual fairness \cite{dwork2012fairness}, causal/counterfactual fairness \cite{kusner2017counterfactual, Zhang2018, chiappa2019path}, group fairness, and intersectional fairness. Individual fairness measures require defining a metric space that encodes differences between individuals \cite{dwork2012fairness} and it is unclear in general how such a metric should be chosen. Causal fairness measures require defining causal relationships between covariates and protected attributes and outcomes, for example in the form of a directed acyclic graph \cite{kusner2017counterfactual}, which is especially challenging to construct in high dimensional settings. In this paper we focus on group fairness definitions. These metrics have the advantage that they can be measured without causal assumptions and without defining a metric over individuals. 

Discrimination criteria can be understood as measuring adherence to one of the following independence statements defined based on a model's predicted score or class, $R$, protected attribute, $S$, and target variable, $Y$: \textit{Independence}:  $R \perp S$, \textit{Separation:} $R \perp S \mid Y$ and \textit{Sufficiency} $Y \perp S \mid R$, \cite{Barocas2016}. We will refer to metrics that require one of these three independence statements above as (statistical) group fairness metrics since independence is observed at the level of the protected groups. Group fairness metrics are usually defined for the classification setting, however many of them naturally lend themselves to regression scenarios or can be adapted simply \cite{steinberg2020fairness}. Intersectional fairness  \cite{buolamwini2018gender} extends group fairness for a more fine-grained, and intersectional, assessment of bias. For example, group fairness measures may assess if a dataset is unbiased across race and gender, whereas an intersectional fairness measures will also assess if the interaction between race and gender are also unbiased.

\subsection{Fairness metrics in survival analysis}
Applications of fairness metrics in clinical decision making have been studied by Pfohl et al. \cite{Pfohl2021-xd}, however this is restricted to the classification setting. Whilst classification metrics may be directly extended to the regression setting, the same does not hold for the more complex survival setting. This is due to: 1) time-to-event datasets including censoring, i.e. patients who are not observed to experience the event of interest; and 2) the prediction of interest is a distribution and not a single value. 

We could find only two strategies evaluating survival fairness in the literature. The first is a transformation of the survival objective, the second is metric based; we briefly discuss each in turn. One approach to evaluating fairness in a survival setting is to assess fairness for binary survival predictions at fixed time points \cite{barda2020developing}, e.g. three years in the future. This strategy can yield viable and fair predictors \textit{if} the selected horizon perfectly coincides with the time-point at which decisions are made. However, this is generally not the case and therefore evaluation of such models usually results in over-confidence in model performance due to `improper' evaluation \cite{Blanche2019}, thus this strategy is not fair. On the other hand, Keya et al \cite{Keya2021-ix} proposed predictive metrics -- individual, group, and intersectional -- for survival fairness that do not require any objective transformation. Whilst this is a great step forward, their metrics are only applicable to linear predictors, such as from the Cox Proportional Hazards (PH) model \cite{Cox1972}. Whilst the Cox PH is arguably the most popular model in survival analysis, this limitation means that their metrics cannot be applied to the majority of machine learning models.

\section{Fairness in survival analysis}

Survival analysis is a task in which one attempts to predict the probability of an event occurring over time. For example, predicting the risk of a patient dying of a disease after diagnosis, predicting when a customer will default on a loan (`duration analysis'), or predicting the probability of a lightbulb failing over time (`reliability analysis'). Survival analysis is distinct from regression as we are interested in making predictions from \textit{censored} time-to-event data. A censored observation is one in which the event of interest does \textit{not} occur. Survival models are fit to estimate the functional relationship between a set of covariates $X$ and time until an event of interest takes place $Y$. We assume, for all observations, that there exists both a hypothetical survival time $Y$ \textit{and} censoring time $C$ (the last recorded time for an observation). We define $T := \text{min}(Y, C)$ as the observed outcome time and $\Delta := \II(Y = T)$ as the survival indicator. Survival models are fit on the survival tuple $(X, T, \Delta)$. For this paper we only consider the right-censoring survival setting. Data is assumed to consist of $n$ observations drawn i.i.d.\ from  a data-generating distribution $\mathrm{P_{x,y}}$. In fairness contexts we further assume there exists one or multiple \textit{sensitive} attributes, $S$, assigning groups to each observation. 

\subsection{Experiment}

\paragraph{Analysis}
We are interested in understanding how well existing losses capture bias in survival datasets. We apply biasing algorithms to 29 published survival datasets (Appendix \ref{app:data}), fit a random survival forest (RSF) \cite{Ishwaran2008} on these biased datasets and then evaluate fairness as $F_L = |L_A - L_D|$ where $F_L$ is the fairness measured by loss $L$, and $L_A, L_D$ are the model performance on the advantaged and disadvantaged subgroups respectively measured by loss $L$. Analysis is performed with \pkg{mlr3proba} \cite{mlr3proba} in \Rstats \cite{rstats}. We fit an RSF as it robustly returns multiple prediction types \cite{Sonabend2022} that can be assessed with calibration, discrimination and scoring rule measures. We increased the proportion, $\sigma$, of biased observations in the disadvantaged data from 0\% to 90\% and asserted that a loss, $L$, could capture bias in our datasets if there was a significant Spearman rank correlation between $F_L$ and $\sigma$, and a significant t-test after regressing $F_L$ on $\sigma$.

\paragraph{Biasing algorithms}
We created biasing methods that mimic identified real-world sources of bias \cite{Mehrabi2019}. Our first method simulates measurement bias by randomly permuting the covariates for an increasing proportion of disadvantaged observations for each dataset, whereas our second method simulates representation bias by increasingly undersampling disadvantaged groups for each dataset. These are fully described in Appendix \ref{app:alg}.

\paragraph{Measures}
We consider a range of calibration, discrimination, and scoring rule measures common in the survival literature. We avoid evaluation bias \cite{Mehrabi2019} by only including (strictly) proper scoring rules as other (improper) scoring rules may not accurately identify a superior model over an inferior one. We consider the right-censored log-likelihood (RCLL) \cite{Avati2018}, reweighted survival Brier score (RSBS) \cite{Graf1999, Sonabend2021}, reweighted integrated survival logloss (RISL) \cite{Graf1999, Sonabend2021}, survival negative log-likelihood  (SNL) \cite{Sonabend2021}, van Houwelingen's alpha (CalA) \cite{VanHouwelingen2000}, D-calibration (CalD) \cite{Haider2020}, Harrell's C ($C_H$) \cite{Harrell1982,Harrell1984}, and Uno's C ($C_U$) \cite{Uno2011}. Scoring rules are standardised against a Kaplan-Meier baseline \cite{Graf1995, Kaplan1958}.

\paragraph{Results}
We now assess how well the metrics recover increased unfairness, controlled by $\sigma$. Regressing $F_L$ on $\sigma$, we find that for the permutation biasing method, $\sigma$ was a significant predictor of $F_L$ for RSBS, RISL, $C_H$, $C_U$, $CalA$. We also find that for RCLL there is a significant correlation between $F_L$ and $\sigma$ however the regression slope is too small to be meaningful. There is also a significant relationship between $F_L$ and $\sigma$ for SNL after applying the undersampling algorithm (Table \ref{tab:lm} and Appendix \ref{app:res}). 

\begin{table}[h] \centering 
  \caption{Let $F_L = \alpha + \sigma\beta$ be our regression model. Table shows intercept, $\alpha$, slope, $\beta$, and Spearman rank correlation ($\rho$) for each of the measures and permutation (left) and undersampling (right) biasing methods. An `\textasteriskcentered' indicates a p-value less than 0.05 after Holm's correction.} 
  \label{tab:lm} 
\begin{tabular}{c|ccc|ccc} 
\toprule
Measure & $\alpha$ & $\beta$ & $\rho$ & $\alpha$ & $\beta$ & $\rho$ \\
\midrule
& \multicolumn{3}{c|}{Permutation} & \multicolumn{3}{c}{Undersampling} \\
\midrule
RSBS & 0.049 & 0.078$^*$ & 0.976$^*$ & 0.035 & 0.066$^*$ & 0.855$^*$ \\
RISL & 0.045 & 0.063$^*$ & 0.976$^*$ & 0.033 & 0.058$^*$ & 0.891$^*$ \\
SNL & 0.018 & 0.001 & 0.248 & 0.014 & 0.083$^*$ & 1.000$^*$ \\
RCLL & 0.018 & 0.009 & 0.879$^*$ & 0.022 & 0.088$^*$ & 1.000$^*$ \\
$C_H$ & 0.024 & 0.129$^*$ & 1.000$^*$ & 0.017 & 0.083$^*$ & 1.000$^*$\\
$C_U$ & 0.031 & 0.124$^*$ & 1.000$^*$ & 0.024 & 0.078$^*$ & 0.976$^*$ \\
CalA & 0.027 & 0.011$^*$ & 0.891$^*$ & -0.015 & 0.197$^*$ & 0.952$^*$ \\
CalD & 2.686 & 0.487 & 0.721$^*$ & 2.861 & 0.646 & 0.612\\
\bottomrule
\end{tabular} 
\end{table} 

\subsection{Discussion}
There is a significant relationship between $\sigma$ and $F_L$ for both measures of separation, $C_H$ and $C_U$. This is intuitive as the biasing methods prevent the model learning the true risk for disadvantaged observations and therefore cannot estimate the difference in risk between advantaged and disadvantaged observations. Secondly, $CalA$ does detect bias in the data whereas $CalD$ does not. $CalA$ evaluates if a model correctly predicts the number of events in the test set. In contrast, $CalD$ evaluates if the predicted survival functions are distributed according to $U(0,1)$. The significant result with $CalA$ indicates that the model cannot predict the number of events in the disadvantaged groups from either biasing method. Whereas the results with $CalD$ may indicate a more complex relationship between distributional calibration and fairness, which has already been demonstrated in the classification setting \cite{Pleiss2017}. Of the scoring rules, only RSBS and RISL could detect the bias from both methods with a regression slope of a meaningful magnitude. This is a promising finding as prior research has demonstrated the usefulness of scoring rules in evaluating fairness \cite{Glymour2019}.

\section{Conclusions}
Algorithmic fairness is an important concept to assess how much bias is present in datasets and subsequently picked up by models. This is especially important in survival analysis, which often overlaps with areas that requires strong ethical consideration. Despite this, the literature around survival fairness is in its infancy. In this paper we have performed a simple experiment to demonstrate how existing survival measures can be utilised to audit bias in algorithmic fairness. Measures of discrimination appear to be optimal for capturing bias however these do not paint a full picture as they ignore model calibration. We have found that the standardised scoring rules, RSBS and RISL, are interpretable, capture both calibration and discrimination, and can detect the bias from our algorithms. We believe future work should consider more complex biasing methods including temporal methods that introduce bias after a certain time-point. Finally, whilst predictive metrics have been proposed for risk predictions \cite{Keya2021-ix}, these have yet to be reviewed in the literature and there is also potential to extend this work to survival distribution predictions. Our paper should help raise awareness that current methods of measuring fairness in survival are very limited and should stimulate interest and exploration in further development.

\paragraph{Disclaimer}
This paper was prepared for informational purposes by the Artificial Intelligence Research group of JPMorgan Chase \& Co. and its affiliates (``JP Morgan’’), and is not a product of the Research Department of JP Morgan. JP Morgan makes no representation and warranty whatsoever and disclaims all liability, for the completeness, accuracy or reliability of the information contained herein. This document is not intended as investment research or investment advice, or a recommendation, offer or solicitation for the purchase or sale of any security, financial instrument, financial product or service, or to be used in any way for evaluating the merits of participating in any transaction, and shall not constitute a solicitation under any jurisdiction or to any person, if such solicitation under such jurisdiction or to such person would be unlawful.

\bibliographystyle{ACM-Reference-Format}
\bibliography{bibliography}

\appendix

\section{Biasing algorithm}
\label{app:alg}

We have a generic biasing algorithm (Algorithm \ref{alg:bias}) that is then specialised with two biasing methods. The algorithm splits the dataset into two equal-sized smaller datasets, one to add bias to ($D_B$) and one to leave untouched ($D_U$). Then we further split $D_B$ into two more datasets, $D_{BA}, D_{BD}$, such that $\sigma\%$ of $D_B$ are in $D_{BD}$. The specific biasing algorithm, $G$, is then applied to $D_{BD}$. $D_{BA}$ and $D_{BD}$ are recombined and then three-fold cross-validation is utilised to evaluate model performance on $D_B$ and $D_U$ and fairness is computed. By splitting the data in this way we are able to: 1) ensure that the proportion of bias is simply controlled; 2) ensure that the model performance does not deteriorate for the unbiased dataset, i.e. if we do not split the original dataset in $D_B$ and $D_U$ then model performance will deteriorate overall due to the bias added to the disadvantaged group, whereas by splitting the dataset we are able to clearly distinguish between the `normal' model performance (when no bias has been added), and the reduced performance due to the added bias.

\paragraph{Biasing method 1 - Permutation}
For the first method we artifically add bias by randomly permuting the covariates of disadvantaged observations. This breaks the relationship between the covariates and outcomes and mimics the real-world problem of data being of lower-quality for disadvantaged groups of people.

\paragraph{Biasing method 2 - Undersampling}
For the second method we artifically add bias by undersampling disadvantaged observations. In Algorithm \ref{alg:bias} this amounts to deleting all observations in $D_{BD}$. This mimics the real-world problem of not capturing enough data for disadvantaged groups of people.

\begin{algorithm}
\caption{Input: $D = (X,T,\Delta)$, survival dataset. $\sigma$, proportion of disadvantaged observations to bias. $L$, survival loss. $M$, survival model. $G$, biasing algorithm. Output: $F_L$, fairness metric.}\label{alg:bias}.
\begin{algorithmic}[1]
\FOR{$i = 1,...,10$}
\STATE $D_B, D_U \gets$ Randomly split $D$ into equal-sized datasets
\STATE $D_{BA},D_{BD} \gets$ Randomly split $D_B$ w.p. $\sigma$
\STATE Apply $G$ to $D_{BD}$
\STATE $D_B \gets (D_{BA}, D_{BD})$
\STATE $L_B \gets$ 3-fold CV of $D_B$ with model $M$ and loss $L$
\STATE $L_U \gets$ 3-fold CV of $D_U$ with model $M$ and loss $L$
\STATE $F_{L;i} \gets |L_B - L_U|$
\ENDFOR
\STATE $F_L \gets \sum_i F_{L;i}/10$
\RETURN $F_L$
\end{algorithmic}
\end{algorithm}

\section{Metric definitions}
\label{app:defs}

 Let $X, Y, C$ be random variables taking values in $\mathcal{X}\subseteq \Reals^n, \mathcal{Y}\subseteq\Reals, \calC\subseteq\Reals$ respectively. In addition, define $T := \min (Y, C)$, and $\Delta := \II(Y = T)$. Finally let $(X_i, Y_i, C_i, T_i, \Delta_i) \stackrel{\text{i.i.d.}}\sim (X, Y, C, T, \Delta)$. We define the following measures and include a brief explanation of their interpretation and use.

\begin{itemize}
    \item Reweighted survival Brier score (RSBS) \cite{Graf1999, Sonabend2021}
    \begin{equation}
    L_{RSBS}(Y, T, \Delta|\KMG) = \frac{\Delta \int_{\calT} (\II(T \leq \tau) - F_Y(\tau))^2 \ d\tau}{\KMG(T)}
    \end{equation}
    This is a strictly proper approximate survival loss that evaluates a survival distribution prediction by measuring the squared distance between the predicted survival probability and whether the event occurs. It is the CRPS with an inverse probability of censoring weighting (IPCW) adjustment.
    \item Reweighted integrated survival logloss (RISL) \cite{Graf1999, Sonabend2021}
    \begin{equation}
    \begin{split}
    &L_{RISL}(Y, T, \Delta|\KMG) = \\ &\frac{\Delta \int_{\calT} \II(T \leq \tau)\log[F_Y(\tau)] + \II(T > \tau)\log[S_Y(\tau)] \ d\tau}{\KMG(T)}
    \end{split}
    \end{equation}
    This is a strictly proper approximate survival loss that evaluates a survival distribution prediction by measuring the logarithm of the predicted survival (or 1-survival) probability and whether the event occurs. It is the integrated loglikelihood with an inverse probability of censoring weighting (IPCW) adjustment
    \item Survival negative log-likelihood  (SNL) \cite{Sonabend2021}
    \begin{equation}
    L_{SNL}(Y, T, \Delta|\KMG) = - \frac{\Delta \log[f_Y(T)]}{\KMG(T)}
    \end{equation}
    This is a strictly proper approximate survival loss that evaluates a survival distribution prediction by taking the logarithm of the predicted probability density function. It is essentially the `usual' negative log-likelihood with an IPCW weighting.
    \item Right-censored log-likelihood (RCLL) \cite{Avati2018}
    \begin{equation}
    L_{RCLL}(Y, T, \Delta) = - \log[\delta f_Y(T) + (1 - \Delta)(S_Y(T))]
    \end{equation}
    This is a strictly proper survival loss that evaluates a survival distribution prediction by taking the logarithm of the predicted probability density function for `dead' observations and the logarithm of the predicted survival function for censored observations.
    \item Harrell's C ($C_H$) \cite{Harrell1982,Harrell1984}
    \begin{equation}
        \begin{split}
        C_H(\phi, T, \Delta) = \frac{\sum_{i\neq j} \II(T_i < T_j, \phi_i > \phi_j)\Delta_i}{\sum_{i\neq j}\II(T_i < T_j)\Delta_i}
        \end{split}
    \end{equation}
    where $\phi$ are predicted risks. \\
    This is a concordance measure that evaluates the discrimination of a ranking prediction by asserting if predicted risks are concordant with observed survival times, i.e., if observation $i$ is predicted to be of higher risk of dying than observation $j$, then this prediction is concordant if $i$ dies before $j$.
    \item Uno's C ($C_U$) \cite{Uno2011}
    \begin{equation}
        \begin{split}
        &C_U(\phi, T, \Delta|\tau) = \\
        &\frac{\sum_{i\neq j} W(T_i)\II(T_i < T_j, \phi_i > \phi_j, T_i < \tau)\Delta_i}{\sum_{i\neq j}W(T_i)\II(T_i < T_j, T_i < \tau)\Delta_i}
        \end{split}
    \end{equation}
    where $\phi$ are predicted risks, $W(t_i) = [\KMG(t_i)]^{-2}$, $\KMG$ is the Kaplan-Meier estimator fit on $(T, 1 - \Delta)$, and $\tau$ is an upper-cutoff timepoint. \\
    This has the same interpretation as Harrell's C but includes an IPCW adjustment to account for censoring.
    \item van Houwelingen's alpha (CalA) \cite{VanHouwelingen2000}
    \begin{equation}
        \alpha(\hat{H}, T, \Delta) = \frac{\sum_i \Delta_i}{\sum_i \hat{H}_i(T_i)}
    \end{equation}
    where $\hat{H}_i$ are individual predicted cumulative hazard functions. \\
    This is a calibration measures that evaluates if a predicted survival distribution is well-calibrated by asserting if the predicted expected number of events is equal (or close to equalling) the true number of observed events.
    \item D-calibration (CalD) -- Algorithm in \cite{Haider2020}. \\
    This is a calibration measures that evaluates if a predicted survival distribution is well-calibrated by asserting if the predicted survival distributions are distributed Uniformly as expected.
\end{itemize}

Let $L_M$ be a scoring rule evaluated on a model $M$ and let $L_K$ be the same scoring rule evaluated on a prediction from the Kaplan-Meier baseline, then the explained residual variation (ERV) \cite{Graf1995} of $L$ is defined as the percentage decrease of $L_M$ from $L_K$:
\begin{equation}
    \tilde{L} = 1 - \frac{L_M}{L_K} = \frac{L_K - L_M}{L_K}
\end{equation}
This allows any scoring rule to be meaningfully interpreted as a percentage increase in performance over a baseline model. We standardise all the scoring rules above (RSVS, RISL, SNL, RCLL) with this method.

\clearpage
\section{Full results}
\label{app:res}

The full results of running our experiment are provided below in tabular and graphical forms.

\begin{table}[h] \centering 
  \caption{Measures computed after running the permutation biasing algorithm a random survival forest, arithmetic mean taken over all datasets. $\sigma$ values are represented in the columns such that $\sigma = 0$ means there is no bias in the disadvantaged dataset whereas $\sigma = 0.9$ means 90\% of observations are biased in the disadvantaged dataset. Measures with  `(ERV)' after them are standardised against a Kaplan-Meier baseline (Appendix \ref{app:defs}).} 
\begin{tabular}{c|cccccccccc} 
\toprule
Measure / $\sigma$ & 0 & 0.1 & 0.2 & 0.3 & 0.4 & 0.5 & 0.6 & 0.7 & 0.8 & 0.9 \\ 
\midrule
& \multicolumn{10}{c}{Permutation Biasing Method} \\
\midrule
$C_H$ & 0.034 & 0.036 & 0.046 & 0.054 & 0.073 & 0.089 & 0.102 & 0.118 & 0.131 & 0.138 \\ 
$C_U$ & 0.038 & 0.043 & 0.054 & 0.06 & 0.076 & 0.096 & 0.109 & 0.12 & 0.132 & 0.141 \\ 
CalA & 0.028 & 0.027 & 0.03 & 0.029 & 0.03 & 0.034 & 0.036 & 0.036 & 0.036 & 0.036 \\ 
CalD & 2.679 & 2.676 & 2.647 & 2.961 & 2.832 & 3.062 & 3.154 & 3.014 & 3.027 & 3.005 \\ 
RCLL & 0.019 & 0.02 & 0.017 & 0.022 & 0.022 & 0.025 & 0.024 & 0.025 & 0.025 & 0.027 \\ 
RISL & 0.049 & 0.049 & 0.054 & 0.064 & 0.065 & 0.085 & 0.085 & 0.091 & 0.098 & 0.096 \\ 
RSBS & 0.054 & 0.054 & 0.061 & 0.072 & 0.075 & 0.098 & 0.096 & 0.105 & 0.116 & 0.113 \\ 
SNL & 0.017 & 0.018 & 0.017 & 0.02 & 0.017 & 0.019 & 0.017 & 0.019 & 0.018 & 0.019 \\ \midrule
& \multicolumn{10}{c}{Undersampling Biasing Method} \\
\midrule
$C_H$ & 0.032 & 0.034 & 0.034 & 0.037 & 0.039 & 0.04 & 0.047 & 0.07 & 0.09 & 0.116 \\ 
$C_U$ & 0.04 & 0.037 & 0.04 & 0.044 & 0.045 & 0.044 & 0.055 & 0.073 & 0.094 & 0.116 \\ 
CalA & 0.028 & 0.028 & 0.027 & 0.033 & 0.041 & 0.043 & 0.055 & 0.079 & 0.133 & 0.27 \\ 
CalD & 2.853 & 2.957 & 3.049 & 3.363 & 3.066 & 2.928 & 3.041 & 3.042 & 3.409 & 3.811 \\ 
RCLL & 0.026 & 0.03 & 0.039 & 0.045 & 0.055 & 0.063 & 0.075 & 0.086 & 0.093 & 0.101 \\ 
RISL & 0.042 & 0.049 & 0.046 & 0.043 & 0.051 & 0.048 & 0.055 & 0.07 & 0.085 & 0.104 \\ 
RSBS & 0.047 & 0.053 & 0.051 & 0.046 & 0.055 & 0.052 & 0.061 & 0.079 & 0.091 & 0.117 \\ 
SNL & 0.017 & 0.022 & 0.03 & 0.037 & 0.045 & 0.055 & 0.066 & 0.075 & 0.081 & 0.086 \\ \bottomrule
\end{tabular} 
\end{table} 
\clearpage
\begin{figure}[h]
    \centering
    \includegraphics[width = 8cm]{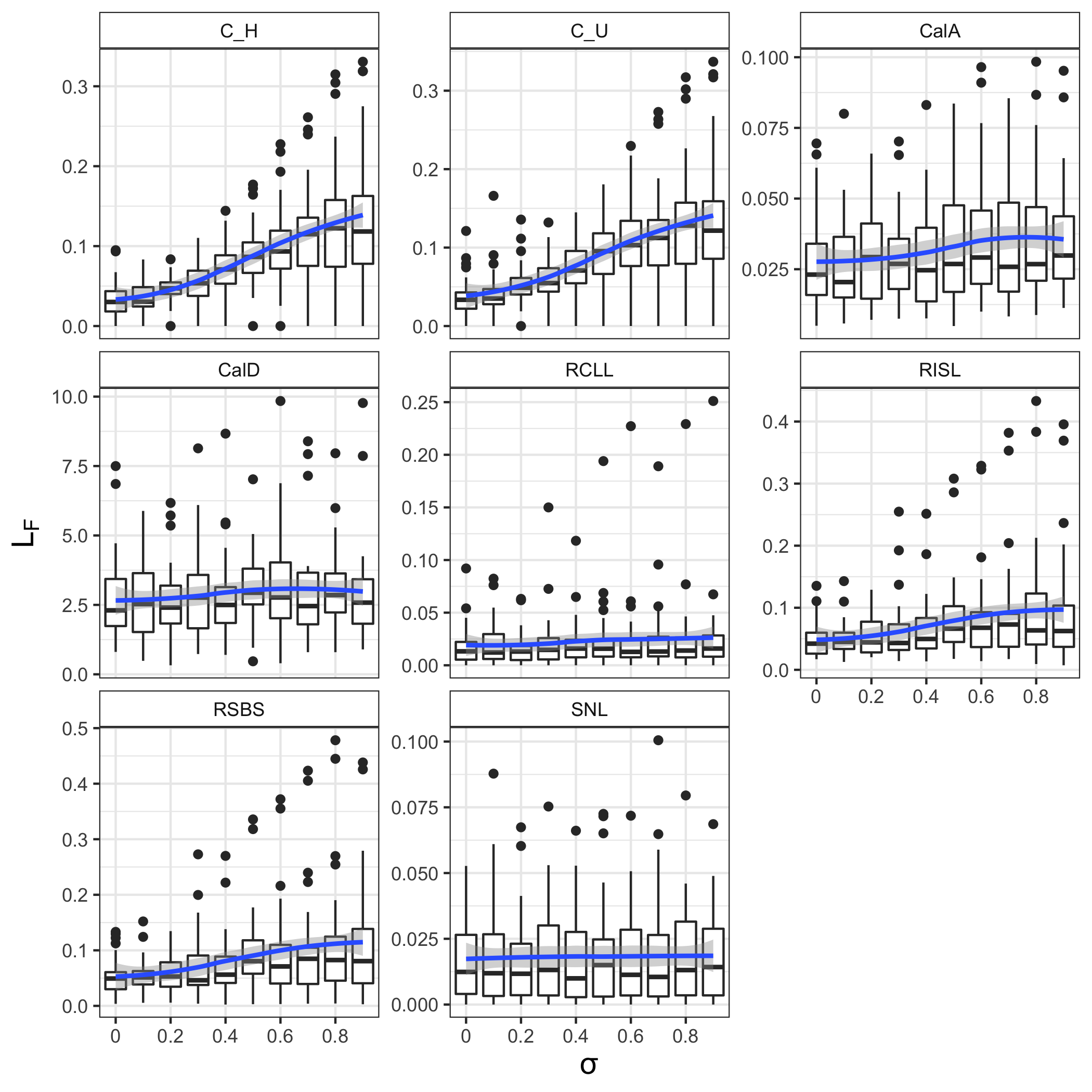}
    \caption{Boxplots of $\sigma$ against $F_L$ over the 29 datasets with the permutation biasing method applied. Blue lines are fit with local polynomial regression.} 
    \Description{Eight boxplots with local polynomial regression smoothing showing the results of the regression of $\sigma$ against $F_L$ over the datasets biased with the permutation method. The boxplots show: $C_H$, $C_U$, $CalA$, $CalD$, $RCLL$, $RISL$, $RSBS$, $SNL$ with increasing slopes for $C_H$, $C_U$, $CalA$, $CalD$, $RISL$ and $RSBS$.}
\end{figure}

\begin{figure}[h]
    \centering
    \includegraphics[width = 8cm]{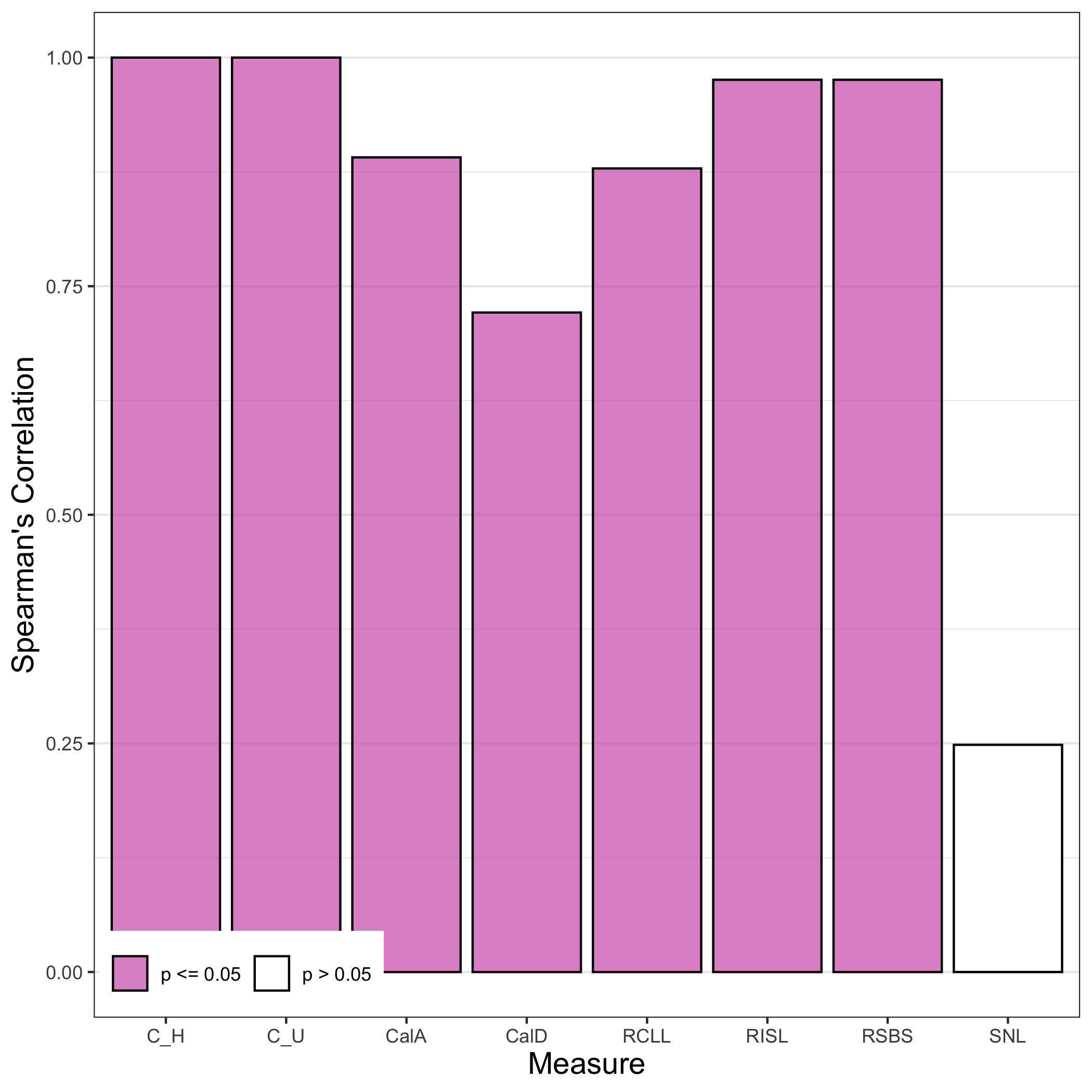}
     \caption{Spearman rank correlation of $\sigma$ against $F_L$ over the 29 datasets with the permutation biasing method applied. Pink bars indicate correlations with $p<0.05$ after correction by Holm's method.}
    \Description{Eight vertical bars showing Spearman rank correlation of $\sigma$ against $F_L$ for $C_H$, $C_U$, $CalA$, $CalD$, $RCLL$, $RISL$, $RSBS$, $SNL$. All bars are pink except for $SNL$ indicating significant correlations.}
\end{figure}

\begin{figure}[h]
    \centering
    \includegraphics[width = 8cm]{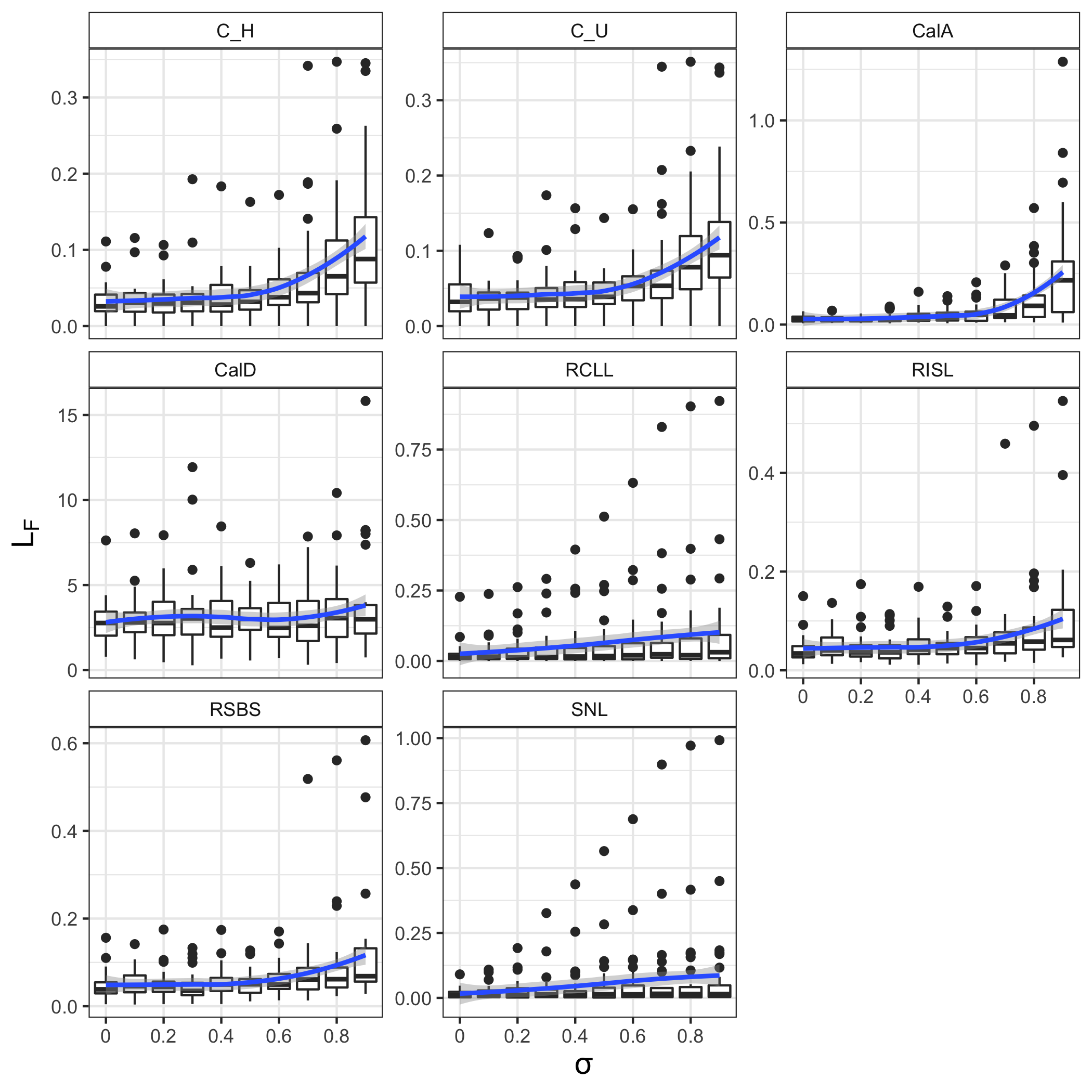}
    \caption{Boxplots of $\sigma$ against $F_L$ over the 29 datasets with the undersampling biasing method applied. Blue lines are fit with local polynomial regression.} 
    \Description{Eight boxplots with local polynomial regression smoothing showing the results of the regression of $\sigma$ against $F_L$ over the datasets biased with the undersampling method. The boxplots show: $C_H$, $C_U$, $CalA$, $CalD$, $RCLL$, $RISL$, $RSBS$, $SNL$ with increasing slopes for $C_H$, $C_U$, $CalA$, $RISL$ and $RSBS$.}
\end{figure}

\begin{figure}[h]
    \centering
    \includegraphics[width = 8cm]{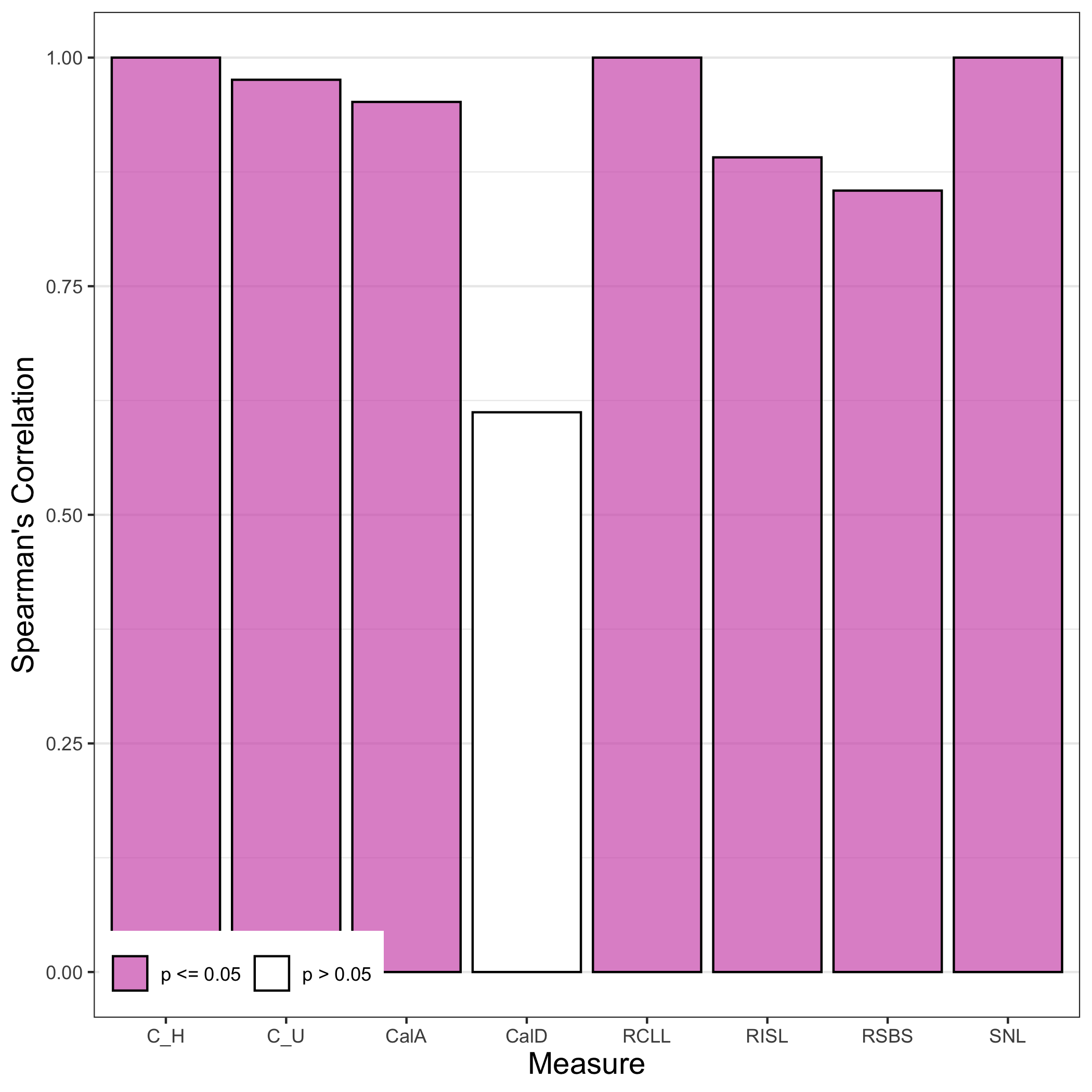}
      \caption{Spearman rank correlation of $\sigma$ against $F_L$ over the 29 datasets with the undersampling biasing method applied. Pink bars indicate correlations with $p<0.05$ after correction by Holm's method.}
    \Description{Eight vertical bars showing Spearman rank correlation of $\sigma$ against $F_L$ for $C_H$, $C_U$, $CalA$, $CalD$, $RCLL$, $RISL$, $RSBS$, $SNL$. All bars are pink except for $CalD$ indicating significant correlations.}
\end{figure}

\clearpage

\section{Datasets}
\label{app:data}

\begin{table}[ht!]
\centering
\caption{Datasets used in experiments. 1. Dataset ID and citation. 2. Proportion of censoring in the dataset, rounded to nearest percentage point. 3-4. Number of continuous and discrete features respectively. 5-6. Total number of observations and features respectively. 7. Number of observed events in dataset. 8. \Rstats package in which the dataset is included.}\label{tab:datasets}
\begin{tabular}{llllllll}
\toprule
\textbf{Dataset}$^1$ & \textbf{Cens \%}$^2$ & $n_C^3$ & $n_D^4$ & \textbf{n}$^5$ & \textbf{p}$^6$ & $n_E^7$ & \textbf{Package}$^8$ \\ 
\hline
aids.id~\cite{dataaidsid} & 60 & 1 & 4 & 467 & 5 & 188 & \pkg{JM}~\cite{pkgjm}  \\
Aids2~\cite{pkgnnet} & 38 & 1 & 3 & 2814 & 4 & 1733 & \pkg{MASS}~\cite{pkgnnet} \\
ALL~\cite{dataall} & 63 & 0 & 4 & 2279 & 4 & 838 & \pkg{dynpred}~\cite{pkgdynpred} \\
bladder0~\cite{databladder0} & 48 & 0 & 3 & 397 & 3 & 206 & \pkg{frailtyHL}~\cite{pkgdynpred} \\
CarpenterFdaData~\cite{Carpenter2002} & 36 & 15 & 11 & 408 & 26 & 262 & \pkg{simPH} \\
channing~\cite{Klein2003} & 62 & 1 & 1 & 458 & 2 & 176 & \pkg{KMsurv} \\
child~\cite{pkgeha} & 79 & 1 & 3 & 26574 & 4 & 5616 & \pkg{eha}~\cite{pkgeha} \\
cost~\cite{datacost} & 22 & 3 & 10 & 518 & 13 & 404 & \pkg{pec}~\cite{pkgpec} \\
e1684~\cite{datae1684} & 31 & 1 & 2 & 284 & 3 & 196 & \pkg{smcure}~\cite{pkgsmcure} \\
flchain~\cite{dataflchain} & 72 & 4 & 3 & 7871 & 7 & 1082 & \pkg{survival} \\
FTR.data~\cite{dataFTRSTR} & 86 & 0 & 2 & 2206 & 2 & 300 & \pkg{MRsurv}~\cite{pkgMRsurv} \\
gbsg~\cite{Katzman2018} & 43 & 3 & 4 & 2232 & 7 & 1267 & \pkg{pycox}~\cite{pkgpycox} \\
grace~\cite{dataapplied} & 68 & 4 & 2 & 1000 & 6 & 324 & \pkg{mlr3proba}~\cite{pkgmlr3proba} \\
hdfail~\cite{pkgfrailtySurv} & 94 & 1 & 4 & 52422 & 5 & 2885 & \pkg{frailtySurv}~\cite{pkgfrailtySurv} \\
kidtran~\cite{Klein2003} & 84 & 1 & 3 & 863 & 4 & 140 & \pkg{KMsurv} \\
liver~\cite{dataliver} & 40 & 1 & 1 & 488 & 2 & 292 & \pkg{joineR}~\cite{pkgjoineR} \\
lung~\cite{datalung} & 28 & 5 & 3 & 167 & 8 & 120 & \pkg{survival} \\ 
metabric~\cite{Katzman2018} & 42 & 5 & 4 & 1903 & 9 & 1103 & \pkg{pycox} \\ 
mgus~\cite{datamgus} & 6 & 6 & 1 & 176 & 7 & 165 & \pkg{survival} \\ 
nafld1~\cite{datanafld1} & 92 & 4 & 1 & 12588 & 5 & 322 & \pkg{survival} \\ 
nwtco~\cite{datanwtco} & 86 & 1 & 2 & 4028 & 3 & 571 & \pkg{survival} \\
ova~\cite{dataova} & 26 & 1 & 4 & 358 & 5 & 266 & \pkg{dynpred} \\
patient~\cite{datapatient} & 79 & 2 & 5 & 1985 & 7 & 416 & \pkg{pammtools}~\cite{pkgpammtools} \\
rdata~\cite{pkgrelsurv} & 47 & 1 & 3 & 1040 & 4 & 547 & \pkg{relsurv}~\cite{pkgrelsurv} \\
reconstitution~\cite{datareconstitution} & 19 & 0 & 2 & 200 & 2 & 162 & \pkg{parfm} \\
std~\cite{Klein2003} & 60 & 3 & 18 & 877 & 21 & 347 & \pkg{KMsurv} \\
STR.data~\cite{dataFTRSTR} & 82 & 0 & 4 & 546 & 4 & 101 & \pkg{MRsurv} \\
support~\cite{Katzman2018} & 32 & 10 & 4 & 8873 & 14 & 2705 & \pkg{pycox} \\ 
tumor~\cite{pkgpammtools} & 52 & 1 & 6 & 776 & 7 & 375 & \pkg{pammtools} \\
uis~\cite{datauis} & 19 & 7 & 5 & 575 & 12 & 464 & \pkg{quantreg}~\cite{pkgquantreg} \\
veteran~\cite{Kalbfleisch2011} & 7 & 3 & 3 & 137 & 6 & 128 & \pkg{survival} \\
wbc1~\cite{datawbc} & 43 & 2 & 0 & 190 & 4 & 109 & \pkg{dynpred} \\
whas~\cite{dataapplied} & 48 & 3 & 6 & 481 & 9 & 249 & \pkg{mlr3proba} \\
\bottomrule
\end{tabular}
\end{table}

\end{document}